\documentclass[pra,twocolumn,superscriptaddress,preprintnumbers,amsmath,amssymb,floatfix]{revtex4}

\usepackage{graphicx}
\usepackage{dcolumn}
\usepackage{bm}
\usepackage{latexsym,epsfig}
\usepackage{textcomp}
\usepackage[breaklinks=true, colorlinks, citecolor=blue, linkcolor=blue, urlcolor=blue]{hyperref}
\usepackage{color}
\usepackage{ulem}

\begin{document}
 \title{Population imbalance in the extended Fermi-Hubbard model}
 \author{A. Dhar}
 \author{J. J. Kinnunen}
 \author{ P. T\"{o}rm\"{a}}
 \email{paivi.torma@aalto.fi}
 
 \affiliation{COMP Center of Excellence, Department of Applied Physics, Aalto University, AALTO, Finland}

 \begin{abstract}
  We study the interplay between population imbalance in a two-component fermionic system and nearest-neighbor 
  interaction using matrix product states method. Our analysis reveals 
  the existence of a new type of Fulde-Ferrell-Larkin-Ovchinnikov phase in the presence of competing interactions. Furthermore, we find distinct evidence for 
  the presence of hidden order in the system. We present an effective model to understand the emergent oscillations in the string correlations 
  due to the imbalance, and show how they can become an efficient tool to 
  investigate systems with imbalance. 
 \end{abstract}
 
 \maketitle

 Imbalance in the different intrinsic spin components lies at the core 
 of systems with finite magnetic moment. It can lead to a variety of novel quantum phases such as the Fulde-Ferrell-Larkin-Ovchinnikov (FFLO) phase, 
 breached pair (BP) states, fully paired and partially polarized phases, phase separated ferromagnetic regions~\cite{paiviPRL2010, sarmaPRA2008, orsoPRL2007, asgariPRB2015, 
 aokiPRA2011, muellerNature2010} in fermionic systems. On the other hand, different types of interactions can compete resulting in 
 structure formation in nature~\cite{andelmanScience1995, kivelsonPRL1994}. Long-range interactions, in particular, induce a miscellany of phases in fermionic systems such
 as charge density wave (CDW), 
 spin density wave (SDW), bond-order wave (BOW), phase separated (PS), singlet superfluid (SSF) and triplet superfluid (TSF) phases~\cite{voitPRB1992, giamarchiBook,
 zhangPRL2004, nishimotoPRL2007, furusakiPRL2002, noackPRB2009, linarxiv2014}. 
 In addition, it introduces a nontrivial structure in the system, one particularly being the subtle hidden order which is revealed by highly non-local 
 string correlation functions. Studies on such correlation functions has revealed interesting phenomena in both 
 bosonic and fermionic systems with long-range interactions~\cite{altmanPRL2006, roncagliaPRL2012, roncagliaPRB2013}.
 But what happens 
 when both imbalance and long-range interactions, nearest-neighbor in particular, are present in the system? The rich interplay between these two parameters 
 in a one-dimensional two-component fermionic system is the focus of this letter.
 
 Recent advances in experiments using ultracold gases have opened up avenues to simulate two spin-component systems and 
 introduce population imbalance between them~\cite{paiviBook}. The ability to trap and cool fermionic atoms with large magnetic moment has 
 made it possible to simulate systems with long-range interactions with controllable magnitude~\cite{lewensteinJPB2011, zollerChemRev2012, santosChapter2015, pfauChapter2015} and 
 image them through a quantum gas microscope~\cite{zwierleinPRL2015, greinerPRL2015, kuhrNature2015, thywissenArxiv2015, takahashiArxiv2015} developed recently.  
 Our results can thus be readily verified experimentally. In general, the results reveal new phases of matter and increase the understanding 
 of 1D quantum magnetism.
 
 
 A system of ultracold two-component fermions in a lattice with nearest-neighbor interaction can be
 described by the extended Hubbard model (EHM) described by the Hamiltonian 
 
 \begin{eqnarray}
  \hat{H} & =& -t \sum_{\langle i,j \rangle, \sigma}{\hat{c}_{i,\sigma}^{\dagger}\hat{c}_{j, \sigma} } + U\sum_{i}{\hat{n}_{i \uparrow}\hat{n}_{i \downarrow}}  
   +V\sum_{\langle i,j \rangle}{\hat{n}_i \hat{n}_j}
  \label{eqn:EHM}
 \end{eqnarray} 
 where $\hat{c}_{i, \sigma}^{\dagger}$ ($\hat{c}_{i, \sigma}$) creates (destroys) a fermion with spin $\sigma$ at site $i$, 
 $\hat{n}_{i, \uparrow}=\hat{c}_{i, \uparrow}^{\dagger}\hat{c}_{i, \uparrow}$ ($\hat{n}_{i, \downarrow}$) is the number operator for
 $\uparrow$ ($\downarrow$) fermions at site $i$, and 
 $\hat{n}_i=\hat{n}_{i, \uparrow}+\hat{n}_{i, \downarrow}$ counts the total number of particles at site $i$. The first term is due to the hopping between neighboring 
 sites whereas the on-site interaction between different spins is denoted by the second term. The third term takes 
 into account the interaction between fermions on nearest neighboring sites. The above Hamiltonian at half filling 
 has been extensively studied before and a rich phase diagram has been reported to exist, with different 
 correlators characterizing the respective phases~\cite{voitPRB1992, 
 zhangPRL2004, nishimotoPRL2007, furusakiPRL2002, noackPRB2009, linarxiv2014}. 
 
 \begin{figure}
 \begin{center}
  \includegraphics[width =0.90\linewidth, height = 0.5 \linewidth]{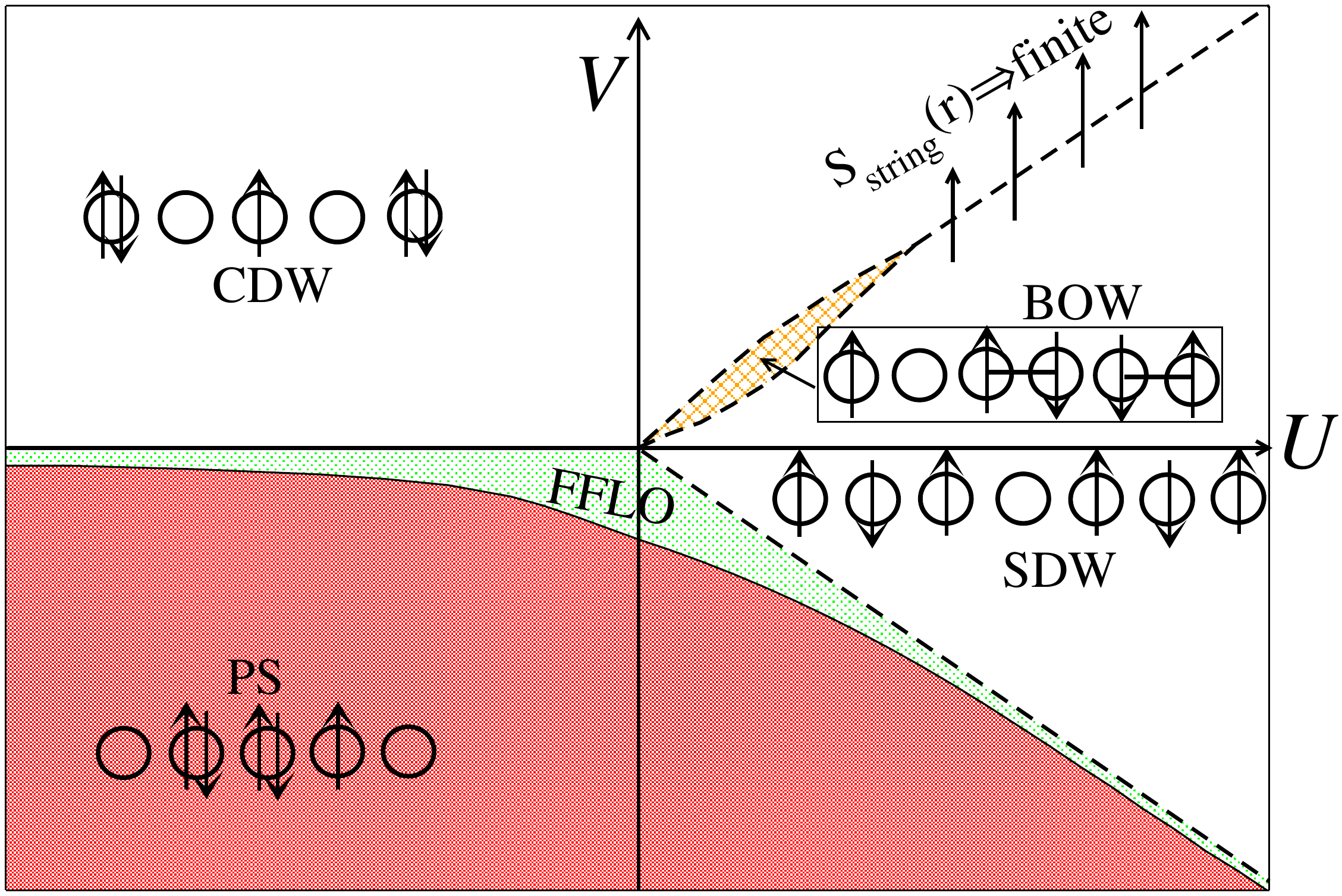}
  \caption{Schematic representation of the different phases for the EHM being modified by the presence of imbalance. It shows the emergence of the FFLO phase
  and finite string correlation found in this letter. }
  \label{fig:phasediag}
 \end{center}
 \end{figure}

 In this work, we proceed in the framework of EHM, but introduce a finite population imbalance between the two 
 components. This is characterized by the polarization defined as $P=\frac{N_{\uparrow}-N_{\downarrow}}{N_{\uparrow}+N_{\downarrow}}$,  
 where $N_{\uparrow}$ and $N_{\downarrow}$ are respectively the total number of up and down spin components in the system. 
 Our primary goal is to look for the emergence of new phases and also to investigate the effects of polarization on the
 different phases existing in the balanced scenario. For this purpose, we select representative points from 
  the various phases extracted from the phase diagrams of EHM in the balanced case reported earlier in the literature~\cite{voitPRB1992, 
 zhangPRL2004, 
 nishimotoPRL2007, furusakiPRL2002, noackPRB2009, linarxiv2014}.

 Using the matrix product states (MPS) method~\cite{schollwockAnP2011, ciracPRL2004}, 
 we simulate a system at zero temperature of size 100 sites (unless otherwise mentioned to lift the degeneracy), with a bond dimension of 500, 
 resulting in an error from the total weight of the discarded states to be 
 less than $10^{-10}$. We fix the filling of spin up particles to $50$ and vary the number of spin
 down particles, thereby achieving various values of polarization, $P$. 
 Such essentially exact calculations of the ground state wave-functions and energies are used to evaluate different correlators needed for our analysis.

\begin{figure}[!t]
  \begin{center}
   \includegraphics[width = 0.95\linewidth]{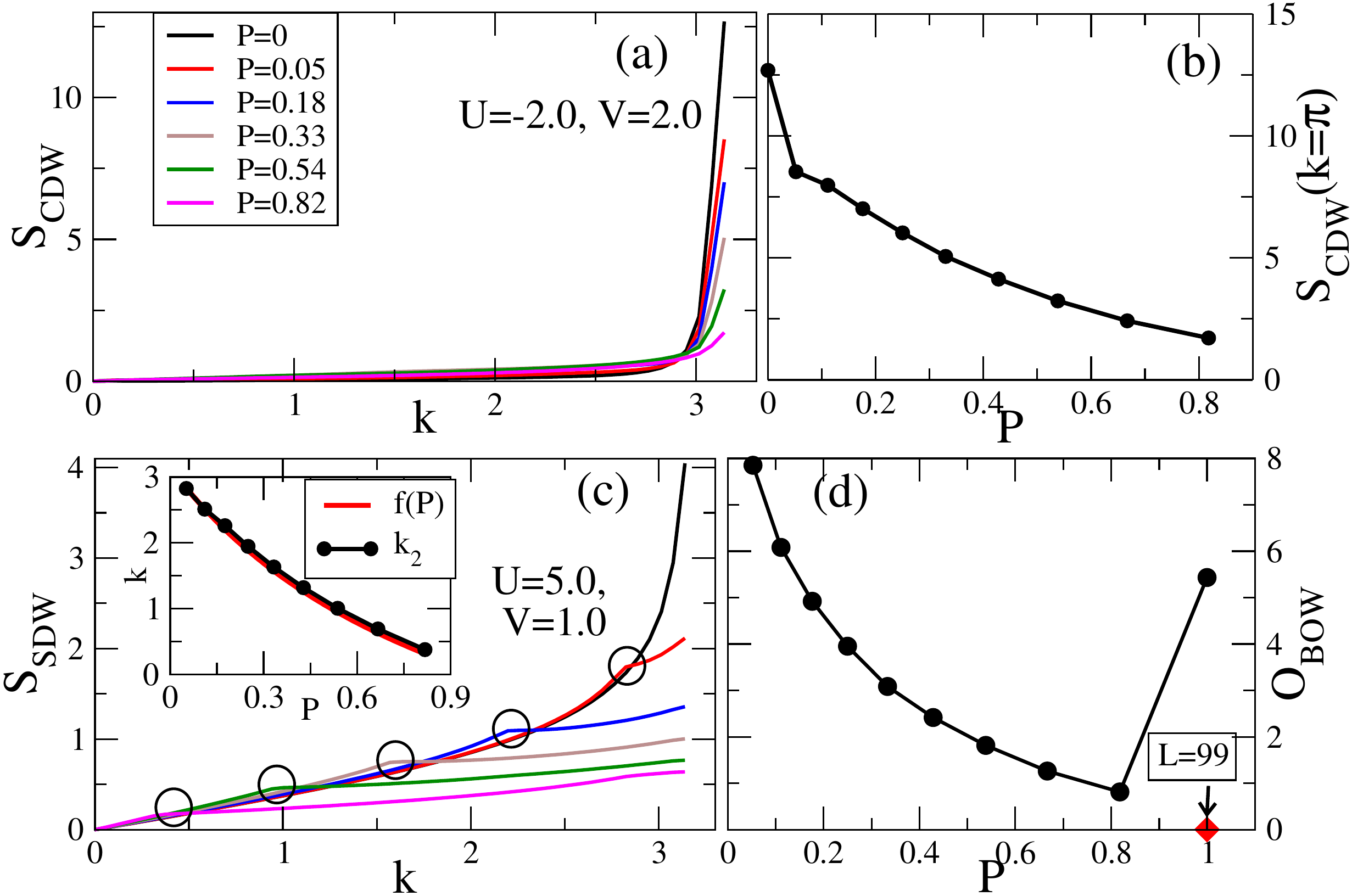}
   \caption{ (a) The density-density structure factor, $S_{\rm{CDW}}(k)$ for different polarizations at $U=-2.0$, $V=2.0$. (b) The monotonic decrease of the peak value of $S_{\rm{CDW}}$
  at $k=\pm \pi$ shown as a function of the polarization P. (c) The spin-spin structure factor $S_{\rm{SDW}}(k)$ for different polarizations at $U=5.0$, $V=1.0$. Black circles
  enclose the positions of the 
  second feature in the presence of imbalance. The color coding for the different polarisation are the same as in (a). Inset shows the momentum values 
  corresponding to the location of the second (bump) as a function of the polarization $P$ along with the function $f(P)= (1-P)/(1+P)$. (d) The bond order wave order parameter
  $O_{\rm{BOW}}$ for different
  polarizations at $U=4.0$, $V=2.1$. The red diamond denotes the value of $O_{\rm{BOW}}$ when degeneracy is lifted by using 99 sites in the simulations. }
  \label{fig:II}
  \end{center}

 \end{figure}

 We begin our analysis by looking at the charge density wave (CDW) phase which appears in an extended region for attractive $U$, repulsive $V$ and also when both $U$ and $V$ are 
 repulsive with $U\lesssim 2V$, have alternate sites as doubly occupied as shown in Fig.~\ref{fig:phasediag}. This phase is characterized by peaks in the structure factor of the density-density correlations
 defined as
 \begin{eqnarray}
 S_{\rm{CDW}}(k)=\sum_{r,r'}{e^{ik(r-r')}(\langle \hat{n}­_{r} \hat{n}_{r'} \rangle - \langle \hat{n}_r \rangle \langle\hat{n}_{r'} \rangle)} .
 \label{eqn:scdw}
 \end{eqnarray}
 For a balanced system, $N_{\uparrow}=N_{\downarrow}$, the structure factor is 
 expected to have peaks at momentum values $k \sim \pm \pi$ in the CDW phase at half filling. Fig.~\ref{fig:II}-(b) indeed shows such well defined 
 peaks of $S_{\rm{CDW}}(k)$. Non-zero values of the polarization $P$ decrease the value of the peaks at $k \sim \pm \pi$, indicating 
 a decrease in the CDW character in the system as $P$ increases. The decrease in the peak value is monotonic 
 in nature as shown in Fig.~\ref{fig:II}-(b). 
 
 The spin-density wave (SDW) phase is characterized by the alternate sites being occupied by up and down spin particles. It 
 shows up in the structure factor of spin-spin correlations, defined as 
 \begin{eqnarray}
 S_{\rm{SDW}}(k)=\sum_{r,r'}{e^{ik(r-r')}(\langle \hat{n}­_{r}^d \hat{n}_{r'}^d \rangle - \langle \hat{n}_r^d \rangle \langle \hat{n}_{r'}^d \rangle)}
 \end{eqnarray} 
 where, $\hat{n}­_{r}^d=\hat{n}_{r, \uparrow}-\hat{n}_{r, \downarrow}$ is the difference between spin-up and spin-down particles at a 
 particular lattice site. In the balanced scenario at half filling, $S_{\rm{SDW}}(k)$ is expected to have peaks at $k \sim \pm \pi$, as observed 
 in Fig.~\ref{fig:II}-(c). With the increase in $P$, we find this peak value to decrease monotonically. 
 In addition, a second feature is observed in the form of a small peak or hump in $S_{\rm{SDW}}(k)$, as denoted by the black circles in 
 Fig.~\ref{fig:II}-(c). This 
 secondary feature can be attributed to the imbalance, and it indicates an additional underlying order appearing 
 in the system. This can be understood if we consider, for example, $50$ up-spin and $25$ down-spin particles. 
 For every down-spin, there are two up-spin particles, resulting in a spin-wave structure with a period of twice as long in the balanced case.
 This translates into momentum of $\pi /2$ 
 for the SDW structure factor. Thus, the momentum values for these secondary features can be 
 approximated by the function $f(P)=\pi N_{\downarrow}/N_{\uparrow}=(1-P)/(1+P)$, which shows excellent agreement with the actual data points 
 as shown in the inset of Fig.~\ref{fig:II}-(c).
 
  The bond order wave (BOW) phase appears in the phase diagram of the EHM for repulsive $U$, $V$ regime as shown in Fig.~\ref{fig:phasediag}. It is characterized by 
  BOW order parameter defined by
 \begin{eqnarray}
 O_{\rm{BOW}}=\sum_{j\sigma}\langle {(-1)^j\left( \hat{c}_{j,\sigma}^{\dagger}\hat{c}_{j+1,\sigma}+h.c. \right) \rangle  }.
 \end{eqnarray}
 Fig.~\ref{fig:II}-(d) shows the monotonic decrease of $O_{\rm{BOW}}$ with the increase of $P$ suggesting 
 the BOW character to decrease with increasing imbalance. This is caused by the smaller number of down-spin 
 particles available to bond with the up-spin particles. Interestingly, the value of $O_{\rm{BOW}}$ in a completely
 polarized case ($N_{\downarrow}=0$) for $L=100$ increases compared to 
 a finite $N_{\downarrow}$. This behavior can be attributed to the presence of large number of degerate ground states when the number of lattice sites is even. 
  The degeneracy is lifted when we use an odd number of lattice sites (L=99) leading to a very small value of the $O_{\rm{BOW}}$ for the fully polarized case, 
  as indicated by the red diamond in ~\ref{fig:II}-(d). 
  
 \begin{figure}
  \begin{center}
   \includegraphics[width=0.95\linewidth]{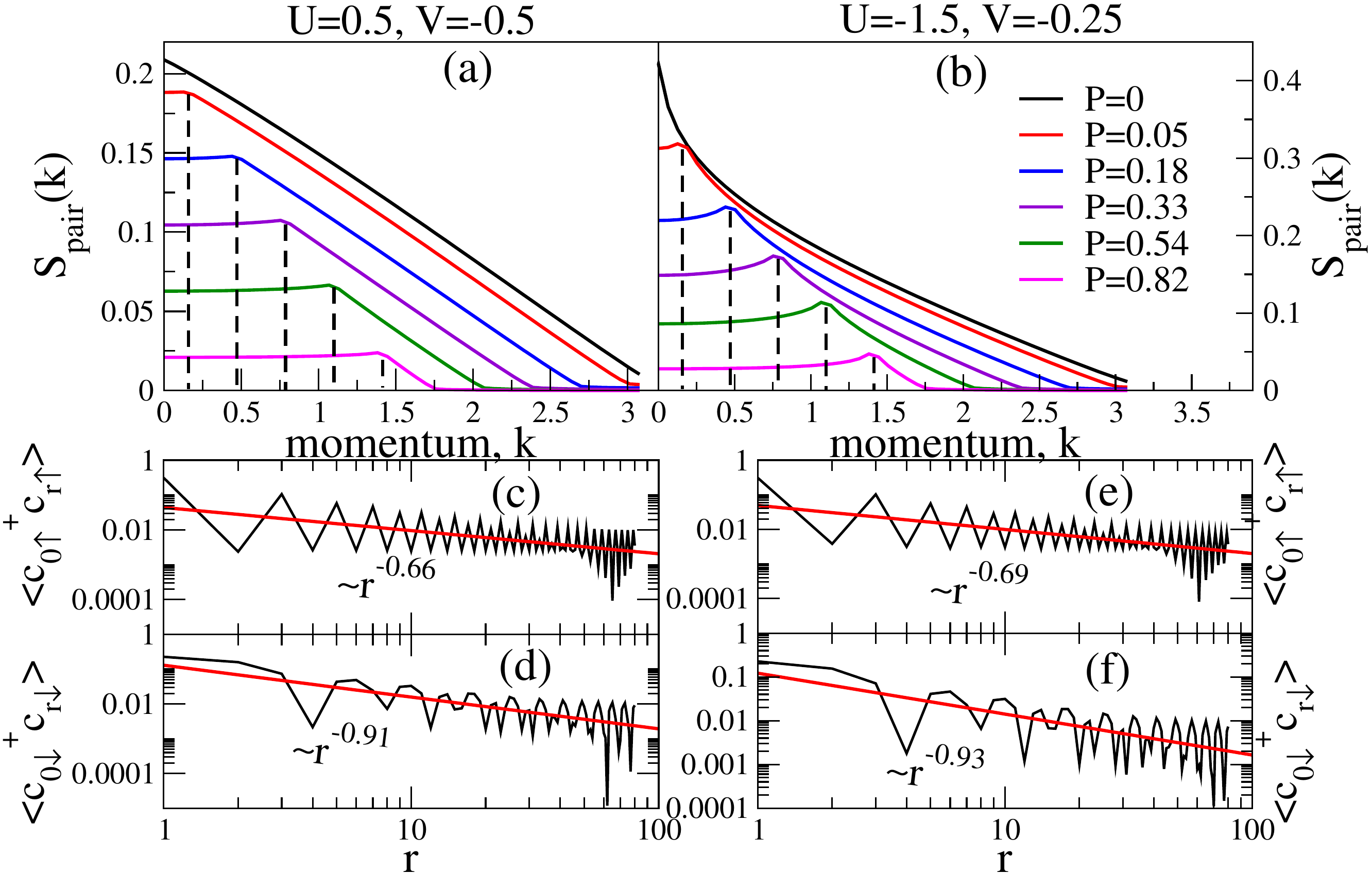}
   \caption{(a) and (b) Plot of the pair momentum distribution, $\rm{S}_{\rm{pair}}(k)$ as a function of momentum $k$ for different values of $P$ for $U=0.5$, $V=-0.5$ 
   and $U=-1.5$, $V=-0.25$ respectively. The dotted lines indicate the values of the FFLO wave vector, $q$ (see text) corresponding to the different values of P.
   Single particle density matrix for spin up [(c), (e)] and down [(d), (f)] particles (black curves) with different values of 
  $P$ for the same set of values of $U,V$ as in [(a), (b)]. The red lines indicate a power law fit to these curves with the exponent mentioned in the respective plots.}
  \label{fig:III}
  \end{center}

 \end{figure}
  
 We now turn our attention to the evaluation of the pair correlation function in the singlet and triplet superfluid (SSF and TSF) phases, defined as
 $\rho_{ij}^{\rm{pair}}=\langle \hat{c}^{\dagger}_{i\uparrow} \hat{c}^{\dagger}_{i\downarrow} \hat{c}_{j\downarrow} \hat{c}_{j\uparrow} \rangle$.
 The pair momentum distribution, $\rm{S}_{\rm{pair}}(k)$, defined as the Fourier transform of $\rho_{ij}^{\rm{pair}}$, indicates the 
 momenta of the Cooper pairs. In the balanced case, the pair momentum has a peak at zero momentum. Switching on 
 the imbalance leads to peaks for non-zero values of the momentum as shown in Figs.~\ref{fig:III}-(a) and (b). Such a behavior signals the appearance of 
 the 1-D analogue of Fulde-Ferrell-Larkin-Ovchinnikov (FFLO) phase [\cite{paiviPRL2010, kajalaPRA2011}, and references therein]. The non-zero momentum 
 values corresponding to the peaks in $S_{\rm{pair}}$ are found to be 
 approximately equal to the FFLO wave-vectors: $q\simeq \pi/[L(N_{\uparrow}- N_{\downarrow})]=\pi \rho P$, where $\rho=(N_{\uparrow}
 + N_{\downarrow})/L$, as shown by the dotted lines in Figs.~\ref{fig:III}-(a) and (c). 
 We also look at the single-particle density matrix of 
 both spin up ($\langle \hat{c}_{0, \uparrow}^{\dagger} \hat{c}_{r, \uparrow} \rangle$) and down ($\langle \hat{c}_{0, \downarrow}^{\dagger} \hat{c}_{r, \downarrow} \rangle$) 
 particles to look for 1-D superfluid (SF) signatures corresponding to these ($U,V$) values.
 Fig.~\ref{fig:III}-[(c) to (f)] shows the existence of long-range correlation, implying the existence of non-exponentially decaying SF order 
 even in the presence of imbalance. Previous works have predicted the presence of FFLO phase in the population imbalance systems with attractive $U$. In 
 Fig.~\ref{fig:III}-(a), we see signatures of FFLO phase even for repulsive $U$ with $V$ being attractive. Our analysis thus reveals the existence of a new type of 
 the FFLO phase.


\begin{figure*}[htbp]
 \includegraphics[width=0.48\linewidth]{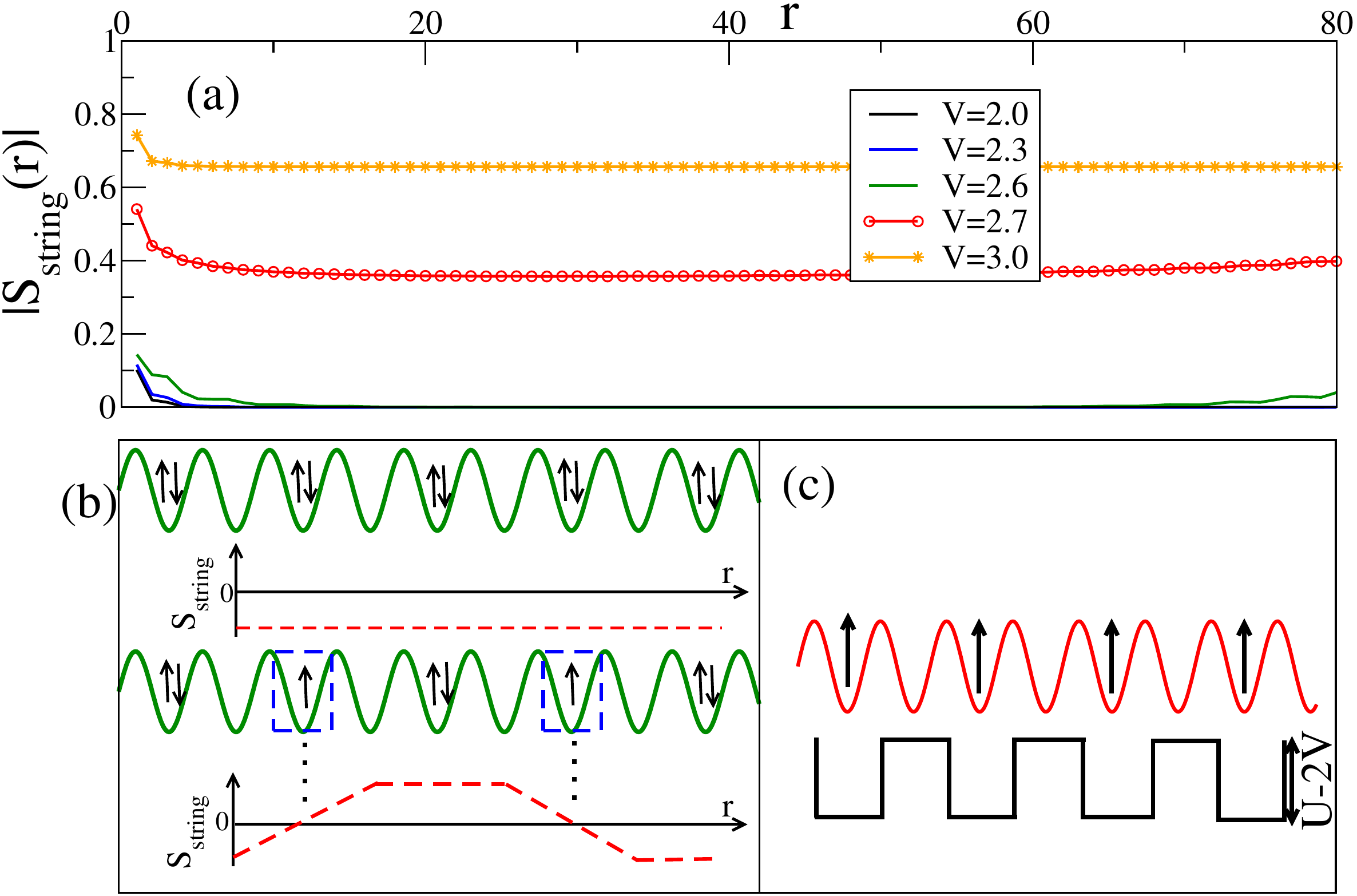}\hspace*{0.5cm}
 \includegraphics[width=0.48\linewidth]{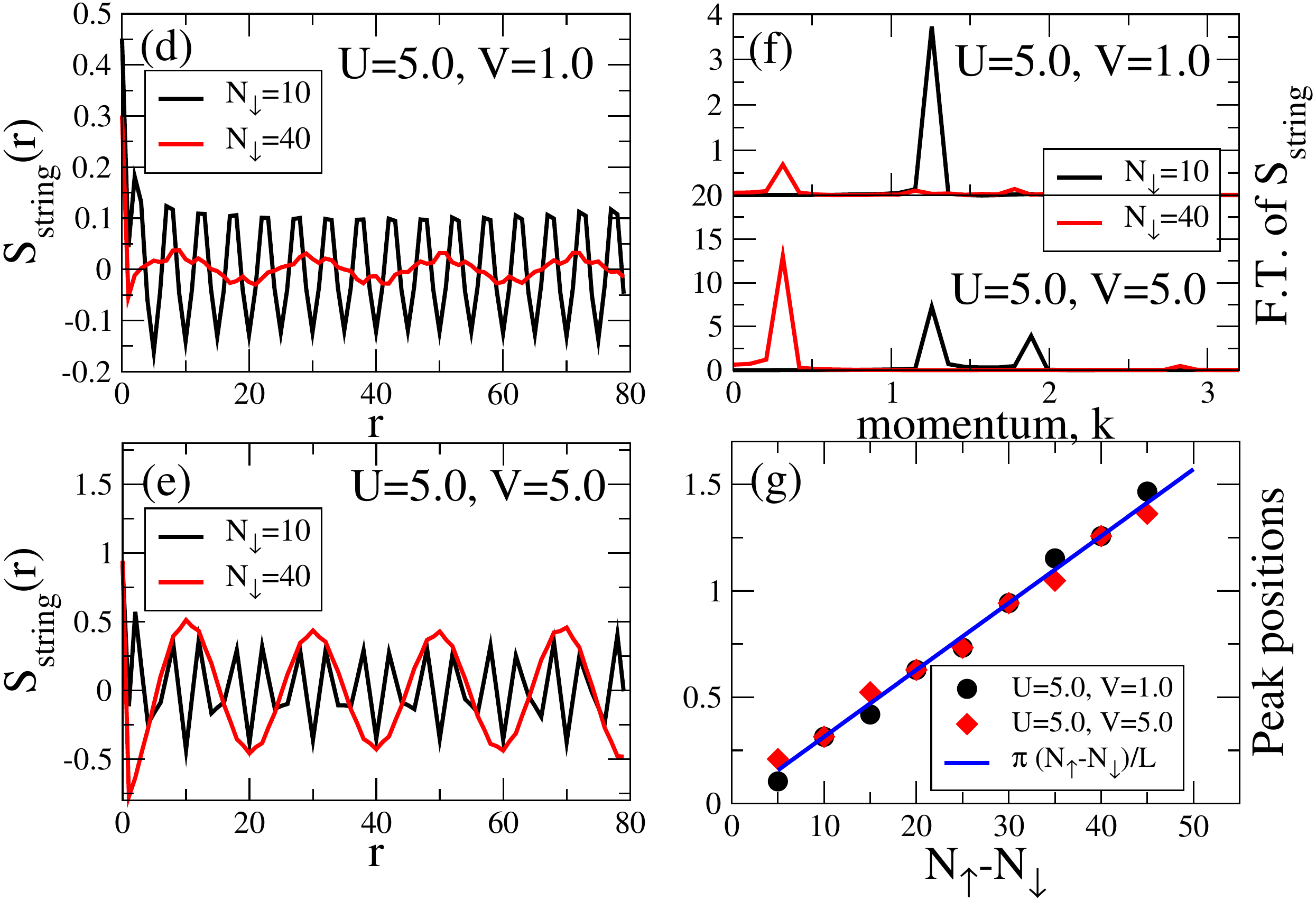}
 \caption{  
 (a) The absolute value of the string correlation with different values of $V$ for $U=5$ in the balanced case, showing the emergence of string order.
  (b)  Schematic diagram to show how string correlation corresponds to the positions of the holes 
 in the $\downarrow$-component. In the balanced 
   case, the string correlation $S_{\rm{string}}$ has a constant value of -1 shown in the top part of the figure. The bottom part shows the
   presence of imbalance ($N_{\downarrow}=N_{\uparrow}/2$),
   resulting in the appearance of holes 
   as denoted by blue dashed boxes. The string correlator has nodes corresponding to the positions of the holes as indicated by the black dotted 
   lines. (c) An effective model of a series of box potentials as seen by a single 
 $\downarrow$-component/hole for strong interactions.
  (d) and (e) Plot of string correlation function for two values of polarization with $U=5$, $V=1$ and $U=5$, $V=5$ respectively. (f) shows the corresponding 
  fast Fourier transforms done
 on the data to extract the frequencies. (g) Primary peak positions of the Fourier transforms of the string correlations 
 as a function of spin imbalance along with the function $\pi \times(N_\uparrow - N_\downarrow)/L$. }
 \label{fig:IV}
\end{figure*}
 
 Motivated by recent reports of hidden orders generating gaps in EHM~\cite{ roncagliaPRL2012, roncagliaPRB2013}, we look for it in the region where both $U$ and $V$ are repulsive. 
Earlier works on the extended Bose-Hubbard 
 model with bosons in an optical lattice~\cite{altmanPRL2006, rossiniNJP2012} had found long range string correlations characterising the Haldane insulator phase. 
 We define the relevant string correlation function as
 
 \begin{eqnarray}
  S_{\rm{string}}(|r-r'|)= \langle \delta \hat{n}_r e^{i \pi \sum_{k=r+1}^{r'-1} \delta \hat{n}_k} \delta \hat{n}_{r'} \rangle
 \end{eqnarray}
 where $\delta \hat{n}_i=\hat{n}_{i, \uparrow}+\hat{n}_{i, \downarrow}-1$. Fig.~\ref{fig:IV}(a) shows the emergence of long-range string correlations
 as V is increased for a fixed large value of U. An odd number of sites ($L = 99$) is used to remove the
degeneracy in the ground state. 
For larger values of $V$, the string correlation is seen to approach a finite value asymptotically, compared to zero for lower values of $V$, 
with the transition occuring close to the SDW-CDW transition at $U=2V$.
Spin-imbalance creates oscillations in the string order function but, even more importantly, it
induces also a finite string correlation in systems which had zero string order in the balanced scenario.
We find that the origin of these effects are in the wavefunctions of the holes
created in the $\downarrow$-component, as explained below.

To understand the connection between the origin of oscillations in the string correlation and imbalance, we consider a case for which $U<2V$. 
In the homogeneous balanced scenario, the system will exhibit CDW nature 
as shown in the upper part of the Fig.~\ref{fig:IV}(b). 
This results in a finite string correlation function. However, removing some $\downarrow$-atoms (creating holes) in a translationally invariant system will create evenly 
spaced nodes in the string correlation corresponding to the locations of the holes, as indicated by the lower part of Fig.~\ref{fig:IV}(b). 
If there are $N(=N_\uparrow-N_\downarrow)$ equally spaced nodes present in a system of $L$ sites, then the wave-vector of the oscillations (or the momentum values from 
the Fourier transform) will be $\pi (N_\uparrow-N_\downarrow)/L$. Fig.~\ref{fig:IV}(f) shows the Fourier spectra of string 
correlations corresponding to Figs.~\ref{fig:IV}(d) and (e) in the presence of imbalance. Fig.~\ref{fig:IV}(g) shows the primary momentum values of the oscillations as a function
of the imbalance, and it 
does show a remarkable conformity with the above scaling. The scaling $\pi (N_\uparrow-N_\downarrow)/L$ was derived here assuming 
localized holes as in Fig.~\ref{fig:IV}(b) lower part. However, localized holes would result in  broad Fourier spectra, unlike
the narrow peaks observed in Figs.~\ref{fig:IV}(f).

In order to go beyond localized holes, we now present an effective single-component model.
Due to nearest neighbour repulsion between half-filled lattice of $\uparrow$-atoms, the $\uparrow$-atoms arrange themselves in a crystal order with alternating occupied/empty-sites. 
Since the interaction is strong, the $\uparrow$-atoms are well localized, and consequently the $\downarrow$-atoms feel an effective static potential in which 
every odd site has an energy shift $U$ and every even site shift $2V$ . In practice, this translates into an alternating potential with wells
of depth $U-2V$ as shown in Fig.~\ref{fig:IV}(c). This potential landscape results in a two-band structure in the single-particle excitation spectrum
of $\downarrow$-atoms (for details, see Supplementary Material). 
In a balanced system, in which the $\downarrow$-component is half-filled, the lower band is fully filled while the upper band is completely empty.

When spin-imbalance is introduced, the first created hole is in the highest lying state of the lower band. 
This hole has a slowly oscillating probability distribution with wavelength equal to twice the size of the system (for details, see Supplementary Material).
Thus the node that in the above simple picture would have been located at the center of the lattice, is now replaced by a hole that has probability distribution spread over 
the whole lattice, but the maximum is still at the center. This smooth spreading with the wavelength of twice the system 
size is intuitive as the behaviour of a particle in a box (combined with the fast every-second-site modulation coming from the effective potential of Fig.~\ref{fig:IV}(c)).
Increasing the spin-imbalance will create more holes with faster oscillations, resulting in the Fourier spectra seen in Fig.~\ref{fig:IV}(f).
In addition to the dominant oscillations, the hole wavefunctions will also have contributions from states with finite 
probabilities of finding the particle in the effective barriers depicted in Fig.~\ref{fig:IV}(c). These manifest as rapidly oscillating Fourier
components, seen also as minor secondary peaks at high momenta (larger than $\pi/2$) in Fig.~\ref{fig:IV}(f).

The string correlator is thus sensitive to correlations between doublons and/or empty sites.
Spin-imbalance produces interesting effects, as it involves creating more empty sites in the SDW-phase or removing doublons in the CDW-state.
In a homogeneous system it shares many properties with the CDW and SDW correlators. 
However, the phase factor $e^{i\pi \Sigma_k \delta \hat{n}_k}$ in the definition of $S_{\rm{string}}$, instead of the factor $e^{ikr}$ in a Fourier transform,
allows more flexibility for the correlations between doublons and/or empty sites. This makes it possible to detect orderings that would otherwise remain hidden.


 To conclude, we have studied a system of two-component fermions in the presence of nearest-neighbor interactions described by the 
 EHM. The introduction of spin imbalance has interesting effects on the various phases of the EHM. The CDW, SDW and BOW
 phases decrease in their respective orders with the increase in polarisation. An additional order appears in the 
 SDW with the onset of imbalance. In the SSF and TSF phases, we observe 
 formation of pairs with finite momentum, implying the appearance of a new type of FFLO phase. We find an extended region for which 
 the FFLO phase persists in the presence of finite polarization. Finally, we find finite hidden order in the system. Imbalance brings 
 in oscillations to the string correlations, even when there is no long-range string order in the balanced system. 
 Our in-depth analysis shows that the string correlator can be a powerful tool when analyzing hole and doublon correlations in lattice systems.
 Our results show that the interplay of spin population imbalance and long-range interactions leads to rich physics, especially concerning exotic 
 paired phases and hidden order.

 \section*{Acknowledgements}
 
 This work was supported by the Academy of Finland through its Centre of Excellence Programme (2012-2017) and under Project No. 263347, 251748, 284621 and 
 272490, and by the European Research Council (ERC-2013-AdG-340748-CODE). Computing resources were provided by CSC - the Finnish IT Centre for Science 
 and the Aalto Science-IT Project.

\end{document}


\title{Supplementary Material}
\date{}
\maketitle

\section{The effective model for holes in a lattice}
 
\begin{figure}[h!]
 \begin{center}
  \includegraphics[width=0.9 \linewidth]{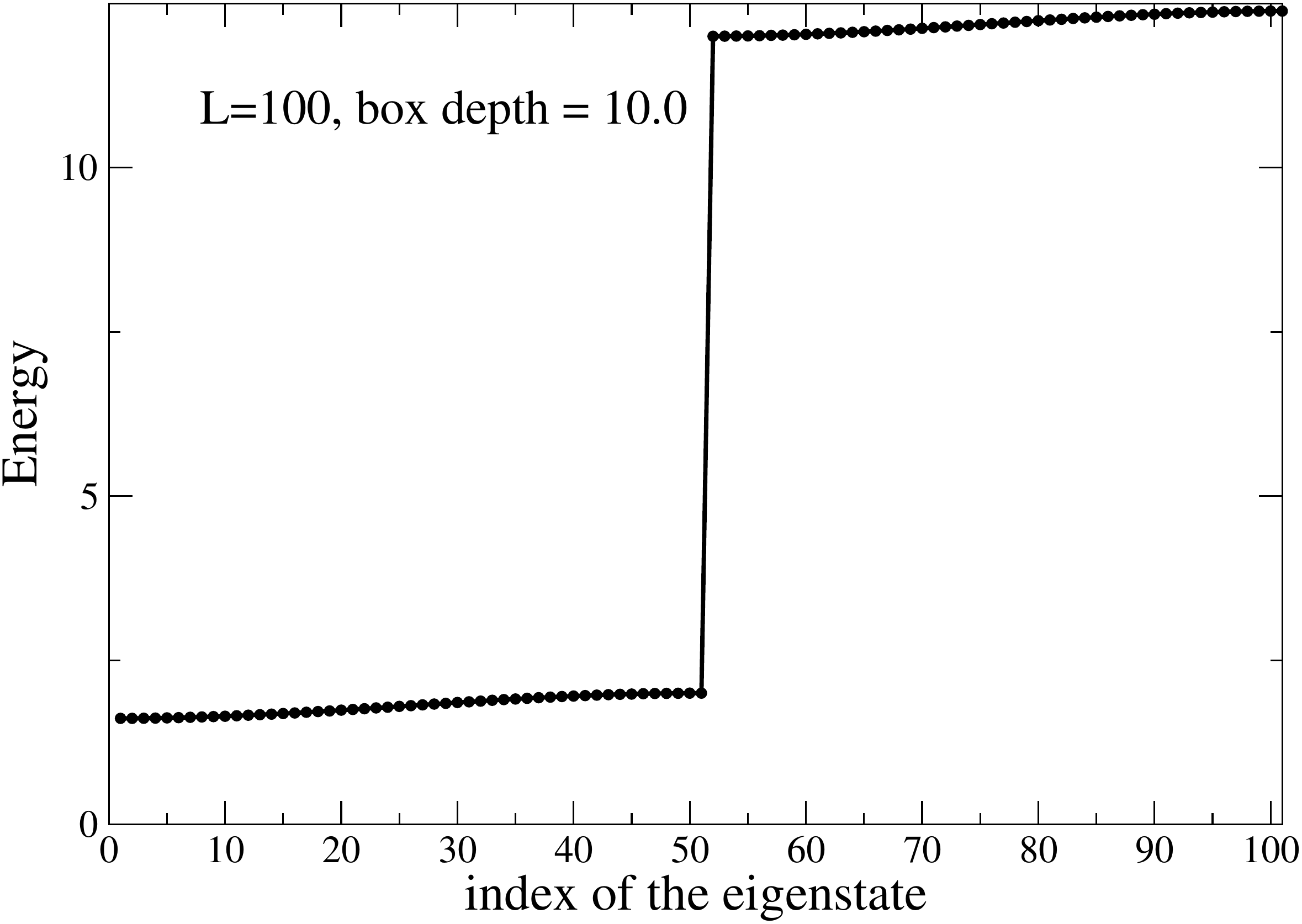}
  \caption{The plot of the energy spectrum from the exact diagonalization code for a single particle in a series of box potentials. It clearly shows 
  two bands with a gap between them.}
  \label{fig:energy}
 \end{center}

\end{figure}

The potential landscape as shown in Fig. 4(c) of the main text results in a two-band structure in the single-particle excitation spectrum of $\downarrow$-atoms
as shown in Fig.~\ref{fig:energy}.
The spectrum has been calculated using an exact diagonalization code. 
Note that the same effective model holds also for the case $U > 2V$, but then sites occupied by the $\uparrow$-atoms become barriers and empty sites become 
wells for the $\downarrow$-atoms.
In a balanced system, in which the $\downarrow$-component is half-filled, the lower band is fully filled while the upper band is completely empty.

\begin{figure}[t]
 \begin{center}
  \includegraphics[width=0.95 \linewidth]{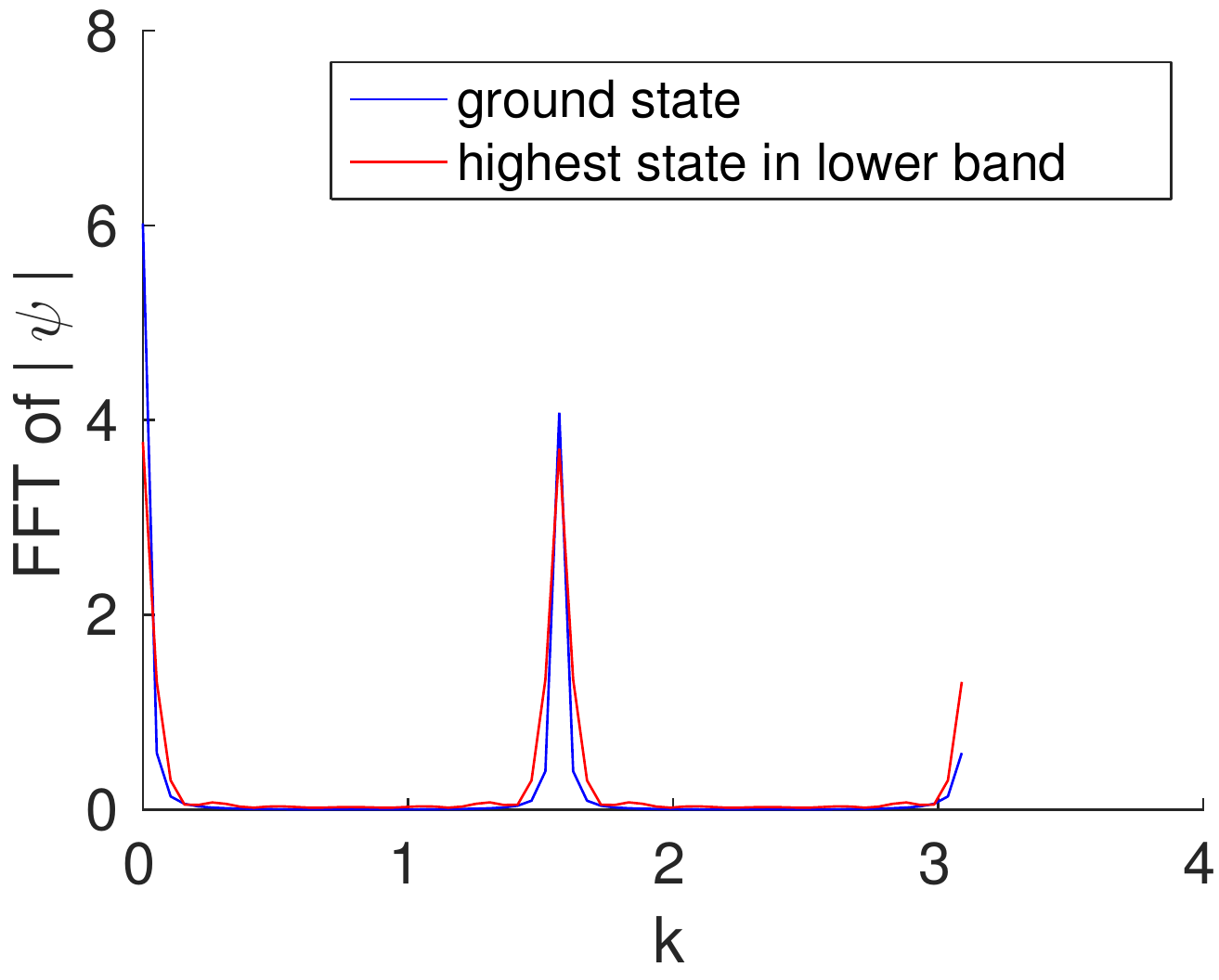}
  \caption{Fast Fourier transform (FFT) of the absolute value of the ground state and the highest state in the lower band (which is the hole wavefunction)
  showing that they have the same frequency component. }
  \label{fig:FFT}
 \end{center}

\end{figure}

When spin-imbalance is introduced, the first created hole is in the highest lying state of the lower band. 
The wavefunction of this hole created will be related to the wavefunction of the highest state in the lower band. A closer analysis (together 
with the particle-hole transformation presented later) shows that this hole wavefunction has the same low frequency component as 
the ground state wavefunction as shown in Fig.~\ref{fig:FFT}. It thus explains the low fequency behavior of the string correlation oscillations for smaller number 
of holes compared to a larger number of holes in the system.

\section{Reducing the model to a single band}

Since the higher band of the two-band model plays no role, we will here consider a simple single-band model. This implies a homogeneous 
lattice without any barriers, thus reducing our system size 
considered in the previous section from 100 to 50. Let us now perform the particle-hole 
transformation in the one-dimensional Hubbard model for non-interacting particles, given by

\begin{eqnarray}
 \hat{H}=-t\sum_{\langle i,j \rangle, \sigma}{\hat{c}_{i, \sigma}^{\dagger} \hat{c}_{j, \sigma} + H.c.}
 \label{eq:hubbard}
\end{eqnarray}

\begin{figure}[t]
 \begin{center}
  \includegraphics[width=0.95 \linewidth]{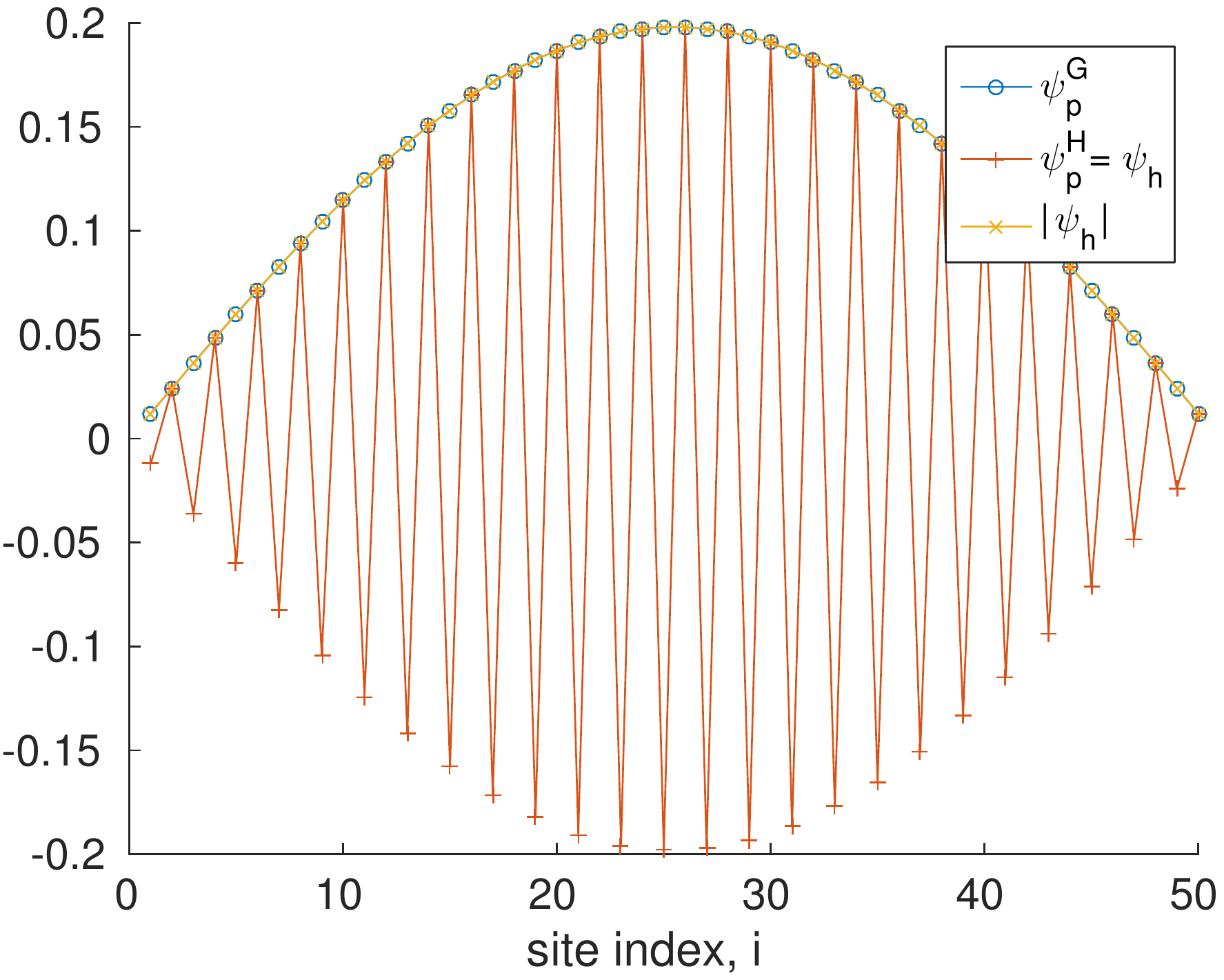}
  \caption{Plot of the wavefunction corresponding to the ground state of Eq.~\ref{eq:hubbard} for particles, $\psi^G_p$ and the highest state for the particles, $\psi^H_p$, which is also 
  equal to the state of the first hole created, $\psi_h$. The absolute value of the hole wavefunction, $|\psi_h|$ is also plotted.
  state is also plotted.}
  \label{fig:wavefunction}
 \end{center}

\end{figure}

We restrict our calculations to a single particle, hence we get rid of the spin index. We first go to the momentum basis to get the 
energy eigenvalues for the particle:

\begin{eqnarray}
 \hat{c}_{k}=\frac{1}{\sqrt{N}}\sum_{l=1}^{N} e^{ikl} \hat{c}_l^{\dagger},
\end{eqnarray}
where $N$ is the number of sites. Such a transformation will diagonalize the Hamiltonian in the momentum basis

\begin{eqnarray}
 \hat{H}=-2t\sum_k{\text{cos}(k) \hat{c}_k^{\dagger} \hat{c}_k}=\sum_k{\epsilon_k \hat{c}_k^{\dagger} \hat{c}_k},
\end{eqnarray}
where $\epsilon_k=-2t\text{cos}(k)$ denotes the dispersion relation, with the allowed momentum values given by $k_n=\pi n/N$, with $1\leq n \leq N$. 
The energy eigenvalues are thus given by $\lambda_n = -2t\text{cos}(n\pi/(N+1))$ and the eigenvectors in the position basis are given
by $V_n = \left( \text{sin}\frac{n\pi}{N+1},  \text{sin}\frac{2n\pi}{N+1}, \cdots , \text{sin}\frac{nN\pi}{N+1} \right)$. The ground state and the highest eigenstate for the particle are shown 
in Fig.~\ref{fig:wavefunction}. Note that the system is finite, so instead of simply a homogeneous density for the ground state, as in an infinite lattice for $k=0$, there is a modulation with 
wavelength of twice the system size. For the highest state in the band, this modulation is the same, but overlayed with the rapid oscillations related to the highest momentum in the 
band. Similar considerations would apply for hole densities in an infinite system.

We now perform the particle-hole transformation :

\begin{eqnarray}
 \hat{c}_i=(-1)^i \hat{h}_i^{\dagger}
\end{eqnarray}
where $h^{\dagger}$ is the creation operators for the holes. Such a transformation also divides the system into a bipartite lattice, with the odd sites getting an extra phase. 
This transformation will lead to a similar Hamiltonian as for the particles 

\begin{eqnarray}
 \hat{H}_{\rm{holes}}=-t\sum_{\langle i,j \rangle, \sigma}{\hat{h}_{i, \sigma}^{\dagger} \hat{h}_{j, \sigma} + H.c.}
\end{eqnarray}
with a similar energy dispersion relation, $\epsilon_k=-2t\text{cos}(k)$. 

Now the hole is created by removing the particle from the highest lying state in the original problem. This is the same as adding one hole in the particle-hole 
transformed system, and the wavefunction of the hole will be similar to that of the ground state in the original problem. But 
because of the bipartite lattice, every alternate site will acquire an extra negative sign, resulting in that the hole wavefunction to be oscillating with the same large wavelength
(smaller momenta) as the ground state wavefunction of the particle but superposed with a rapid sign changing modulation, as shown in Fig.~\ref{fig:wavefunction}. 
The string correlator studied in the main text involves density operators $\hat{n}_{i \sigma} = \hat{c}^{\dagger}_{i \sigma} \hat{c}_{i \sigma}$, and the correlator is 
thus sensitive to the density distributions. Hence the rapid phase modulations in the hole wavefunctions are not observed. This is precisely the reason why we observe 
lower momenta in the Fourier transform of the string correlator when we have lesser number of holes.